\documentclass[11pt,a4paper]{article}

\usepackage[T1]{fontenc}
\usepackage[utf8]{inputenc}
\usepackage{lmodern}
\usepackage{amsmath,amssymb,amsthm,mathtools,bm}
\usepackage{geometry}
\usepackage{booktabs}
\usepackage{hyperref}
\usepackage[nameinlink,noabbrev]{cleveref}
\usepackage{authblk}
\usepackage{tikz}
\usepackage{xcolor}
\usetikzlibrary{arrows.meta,calc}

\hypersetup{colorlinks=true,linkcolor=blue,citecolor=blue,urlcolor=blue}
\numberwithin{equation}{section}

\newcommand{\RR}{\mathbb{R}}

\newcommand{\dd}{\,\mathrm{d}}
\newcommand{\e}{\mathrm{e}}
\newcommand{\ii}{\mathrm{i}}

\newcommand{\hs}{\langle h\rangle}

\newcommand{\bra}[1]{\langle #1|}
\newcommand{\ket}[1]{|#1\rangle}
\newcommand{\ketbra}[2]{|#1\rangle\langle #2|}

\newtheorem{theorem}{Theorem}[section]
\newtheorem{proposition}[theorem]{Proposition}
\newtheorem{corollary}[theorem]{Corollary}
\theoremstyle{remark}
\newtheorem{remark}[theorem]{Remark}

\newcommand{\Sn}{S^n}
\newcommand{\SnR}{S_R^n}
\newcommand{\SnReps}{S_{R,\varepsilon}^n}

\newcommand{\Xeps}{X_\varepsilon}

\newcommand{\Pheps}{\Phi_\varepsilon}

\begin{document}
	
	
	\title{Bound State Analysis of Rank-One Delta Interactions Supported by Deformed Hyperspheres}

	\author[1]{Hilal Demirdöğen}
	\author[1]{Fatih Erman\thanks{Corresponding author. Email: \texttt{fatih.erman@gmail.com}}}
	\author[1]{Merve Nur Sayar}
	
	\affil[1]{Department of Mathematics, \.{I}zmir Institute of Technology, Urla, 35430, \.{I}zmir, Türkiye}
	
	\maketitle

	\begin{abstract}
		We study rank-one delta interactions supported by small normal deformations of
		an \(n\)-dimensional hypersphere \(S_R^n\subset\mathbb{R}^{n+1}\).  The
		interaction is defined by the normalized surface measure on the support and
		therefore corresponds to the rotationally invariant sector of the
		hyperspherical shell problem.  For the undeformed hypersphere, we derive the
		bound-state equation for \(E=-\nu^2\) in terms of the modified Bessel product
		$I_{\frac{n-1}{2}}(\nu R)K_{\frac{n-1}{2}}(\nu R)$.
		We then consider an inward normal deformation
		$X_\varepsilon(\omega)=(R-\varepsilon h(\omega))\omega,\; \omega\in S^n, \; 0<\varepsilon\ll1$.
		Our main result is that, to first order in \(\varepsilon\), the bound-state
		energy depends only on the average normal displacement
		$\langle h\rangle
		=
		\frac1{|S^n|}\int_{S^n}h(\omega)\,d\Omega_n(\omega)$.
		Equivalently, the deformed hypersphere is spectrally equivalent, up to
		\(O(\varepsilon^2)\), to a round hypersphere with effective radius
		\(R-\varepsilon\langle h\rangle\).  In particular, mean-zero deformations do not
		change the rank-one bound-state energy at first order.
	\end{abstract}

	\section{Introduction}
	
	Singular interactions in quantum mechanics provide a useful class of exactly
	solvable models in which the interaction is supported on lower-dimensional
	sets, such as points, curves, surfaces, or hypersurfaces.  The rigorous
	operator-theoretic foundations of such models are presented in the monographs
	\cite{Albeverio,Kurasov2000}, where self-adjoint extension theory and resolvent
	formulas provide a systematic framework for zero-range perturbations.  In this
	setting, Krein-type resolvent formulas play a central role, since they reduce
	the spectral analysis of singular perturbations to the study of a finite- or
	infinite-dimensional principal operator.
	
	Interactions supported on spheres form an important and particularly tractable
	subclass of these models.  A single sphere interaction was studied in detail in
	\cite{AntoineGesztesyShabani1987}, where spherical symmetry and partial-wave
	decomposition reduce the spectral problem to explicit equations in each angular
	momentum channel.  This analysis was extended to finitely many
	\(\delta\)-interactions supported on concentric spheres in \cite{Shabani1988}.
	More general sphere interactions with nonseparated boundary conditions were
	considered in \cite{DabrowskiShabani1988}, and the corresponding scattering
	theory for finitely many concentric sphere interactions was developed in
	\cite{HounkonnouHounkpeShabani1997}.  These works concern local shell-type
	interactions, for which all angular momentum sectors are present.
	
	A broader theory of hypersurface-supported interactions has also been developed.
	Schr\"odinger operators with \(\delta\)- and \(\delta'\)-interactions supported
	on hypersurfaces have been formulated using boundary conditions, extension
	theory, and Krein-type resolvent formulas in
	\cite{BehrndtLangerLotoreichik2013}.  Geometric effects for singular
	interactions supported by curves and surfaces, including curvature-induced bound
	states and strong-coupling asymptotics, were investigated in
	\cite{ExnerKondej2002,ExnerKondej2003,ExnerYoshitomi2002}.  Spectral effects of
	deformed or locally perturbed supports were further studied in
	\cite{ExnerKondejLotoreichik2018}, while spectral comparison results for
	hypersurface interactions were obtained in \cite{LotoreichikRohleder2017}.
	
	The hyperspherical setting is especially relevant for the present work.
	Higher-dimensional delta-shell problems have been studied by radial and
	partial-wave methods in \cite{DemiralpBeker2003,Demiralp2005}.  Hyperspherical
	\(\delta\)-\(\delta'\) interactions were analyzed in
	\cite{MunozCastanedaNietoRomaniega2019}.  Related Lippmann--Schwinger
	formulations for hyperspherical potentials were investigated in
	\cite{PereiraSchmidt2022}, and analytical solutions of the
	Lippmann--Schwinger equation in three-dimensional hyperspherical and
	hyper-pseudospherical spaces were recently obtained in
	\cite{deJesusFortinySchmidt2026}.  These works show that hyperspherical
	geometry provides a natural setting in which singular interactions can be
	treated explicitly.
	
	The present paper is a direct continuation of
	\cite{ErmanSeymenTurgut2022}, where rank-one interactions supported by a circle
	in \(\mathbb{R}^2\) and by a sphere in \(\mathbb{R}^3\), together with their
	small normal deformations, were studied.  In contrast with local shell
	interactions, the rank-one interaction considered here is generated by the
	normalized surface measure on the support.  Consequently, it couples only to the
	constant hyperspherical harmonic, and the spectral problem is reduced to a
	single scalar principal function.  Our main purpose is to extend this rank-one
	deformation analysis to an \(n\)-dimensional hypersphere
	\(S_R^n\subset\mathbb{R}^{n+1}\).  We derive the dimension-dependent principal
	function explicitly, determine the corresponding bound-state equation, and show
	that, to first order in the deformation parameter, the bound-state shift depends
	only on the average normal displacement of the hyperspherical support.

	Delta interactions supported by submanifolds provide a useful class of singular
	perturbations of Schr\"odinger operators.  In the two- and three-dimensional
	cases one may consider interactions supported by a circle or a sphere and
	describe the corresponding rank-one singular perturbation by averaging the wave
	function over the support. We follow the rank-one
	formulation used for the circle and sphere problems: the interaction is not the
	full delta-shell operator containing all angular momentum sectors, but the
	singular rank-one interaction associated with the normalized surface measure.
	Hence the result describes the rotationally invariant, or zero-angular-momentum,
	sector of the usual spherical shell problem. Such singular interactions are formally described by the following Schrödinger operators or Hamiltonians:  
	\begin{eqnarray}
		H & = & H_0-\lambda |\delta_{S^1_R}\rangle \langle \delta_{S^1_R}| \\ 
		H & = & H_0-\lambda |\delta_{S^2_R}\rangle \langle \delta_{S^2_R}|
	\end{eqnarray}
	where $H_0$ is the free part and $S^1_R$ is the circle of radius $R$, $S^2_R$ is the sphere of radius $R$. Let $L(S^1_R)$ be the lenght of the circle of radius $R$ and $A(S^2_R)$ be the area of the sphere of radius $R$. Then, the action of the delta distributions supported by a circle and sphere are defined by
	\begin{eqnarray}
		\langle \delta_{S^1_R}|\psi \rangle & := &  \frac{1}{L(S^1_R)} \int_{S^1_R} \psi \; \dd s \\ 
		\langle \delta_{S^2_R}|\psi \rangle & := &  \frac{1}{A(S^2_R)} \int_{S^2_R} \psi \; \dd \sigma \;.
	\end{eqnarray}
	Here $\dd s$ is the integration element over the circle $S^1$ and $\dd \sigma$ is the integration element over the sphere $S^2$.

	Throughout the paper, we use Dirac's bra--ket notation in the usual distributional sense. This convention is adopted in order to make the presentation accessible to physicists and to keep the connection with the standard resolvent or Green's function (which is the integral kernel of resolvent) formulation transparent. Thus
	\(\ket{\psi}\) denotes a state vector, while \(\bra{\psi}\) denotes the
	corresponding dual vector.  The symbol \(\ket{\delta_{S^{2}_R}}\) (similarly for $\ket{\delta_{S^{1}_R}}$) should not be understood
	as an ordinary square-integrable function, but rather as the distribution
	associated with the normalized surface measure on \(S_R^2\) ($S_R^1$).  
	With this convention, the rank-one operator
	$\ket{\delta_{S_R^2}}\bra{\delta_{S_R^2}}$ acts on a test function \(\psi\) as
	$\ket{\delta_{S_R^2}}\,\langle \delta_{S_R^2}|\psi\rangle$.
	Consequently, the formal Hamiltonian represents a singular attractive interaction supported on the sphere.
	This notation is convenient because the corresponding resolvent takes the
	standard rank-one Krein form \cite{Kurasov2000, ErmanSeymenTurgut2022}:
	\begin{eqnarray}
		R(E)=R_0(E)+ R_0(E) |\delta_{S^2_R} \rangle \frac{1}{\Phi(E)} \langle \delta_{S^2_R}| R_0(E) \;, \label{eq:Kreinsphere}
	\end{eqnarray}
	where $R_0(E)=(H_0-E)^{-1}$ defined on its resolvent set and 
	\begin{eqnarray}
		\Phi(E)= \frac{1}{\lambda}- \langle \delta_{S^2_R}|R_0(E)|\delta_{S^2_R}\rangle \;. \label{eq:principal}
	\end{eqnarray}

	The present work is a continuation of our previous study
	\cite{ErmanSeymenTurgut2022}, where rank-one delta interactions supported by
	circular and spherical geometries were analyzed, and the effect of small normal
	deformations of these supports on the bound-state spectrum was investigated. In
	that setting, it was shown that, to first order in the deformation parameter,
	the change in the bound-state energy has a simple geometric interpretation: the
	deformed support behaves spectrally like the original circle or sphere whose
	radius is shifted by the average normal displacement. The main goal of the
	present paper is to extend this analysis to arbitrary dimension. More precisely,
	we study rank-one delta interactions supported by an \(n\)-dimensional
	hypersphere \(S_R^n\subset\mathbb{R}^{n+1}\), and then deform the support in
	the normal direction. Using the resolvent formulation, we derive the implicit bound-state equation for the
	undeformed hypersphere and compute its first-order variation under small
	normal deformations. This provides a unified \(n\)-dimensional framework that
	contains the deformed circle and deformed sphere results as the special cases
	\(n=1\) and \(n=2\), respectively.

	The paper is organized as follows.  We first fix the notation for the unit
	hypersphere, and its small normal deformation, together
	with the corresponding surface measures.  We then introduce the rank-one
	hyperspherical delta interaction and derive the criteria for
	the undeformed support \(S_R^n\subset\mathbb{R}^{n+1}\) in Section \ref{Sec2:Rank-one hyperspherical delta interaction}. Then,
	we use this result to analyze the existence and uniqueness of the
	bound state and to identify the dimension-dependent critical coupling.  We
	then turn to small normal deformations of the hyperspherical support in Section \ref{Bound State Analysis of Delta Potential Supported on Deformed Hypersphere} and by
	expanding the normalized surface measure and the resolvent matrix element to
	first order in the deformation parameter, we obtain the first-order correction
	to the bound-state equation and to the bound-state energy.  Finally, we discuss
	the geometric interpretation of the result, namely that, to first order, the
	deformed hypersphere is spectrally equivalent to a round hypersphere whose
	radius is shifted by the average normal displacement. A short derivation of plane waves into hyperspherical harmonics is given in Appendix 
	to make the paper self-contained.

	\section*{Notation}
	\label{sec:notation}
	
	We fix the following notation throughout the paper.  The unit hypersphere in
	$\RR^{n+1}$ is
	\begin{equation}
		\Sn=\{\omega\in\RR^{n+1}:|\omega|=1\},
	\end{equation}
	where its parametrization is given by $\omega = \omega(\theta_1,\ldots,\theta_n)$ with 
	\begin{eqnarray}
		\omega_1 & = & 
		\cos\theta_1 \;, \nonumber \\
		\omega_2 & = & \sin\theta_1\cos\theta_2 \;,\nonumber \\
		\omega_3 & = & \sin\theta_1\sin\theta_2\cos\theta_3 \;, \nonumber \\ 
		\vdots \nonumber \\
		\omega_n & = & \sin\theta_1\cdots\sin\theta_{n-1}\cos\theta_n \;, \nonumber \\
		\omega_{n+1}& = & \sin\theta_1\cdots\sin\theta_{n-1}\sin\theta_n \;.
		\label{eq:hyperspherical-omega}
	\end{eqnarray}
	Here $0\leq \theta_1,\ldots,\theta_{n-1}\leq\pi$, $0\leq\theta_n<2\pi$. Thus \(\omega\)
	also encodes the angular direction.
	
	Throughout the paper, \(\omega\in S^n\) denotes a point on the unit sphere
	\(S^n\subset\mathbb{R}^{n+1}\).  Since points of \(S^n\) are unit vectors in the
	ambient space, \(\omega\) will also be regarded as a unit vector and it encodes the angular direction. The standard
	surface element on \(S^n\) is denoted by \(d\Omega_n(\omega)\) and given by
	\begin{equation}
		\dd\Omega_n
		=\prod_{j=1}^{n-1}\left(\sin^{n-j}\theta_j\dd\theta_j\right)\dd\theta_n.
		\label{eq:dOmega-coordinates}
	\end{equation}
	is the standard surface element on the unit hypersphere or intrinsic $n$ dimensional volume measure. This is the generalization of the solid angle element in the two dimensional case $\dd \Omega=\sin \theta \dd \theta \dd \varphi$. The $n$-dimensional surface area of $S^n$ is given by
	\begin{equation}
		|\Sn|:=\int_{\Sn}\dd\Omega_n
		=\frac{2\pi^{(n+1)/2}}{\Gamma\!\left(\frac{n+1}{2}\right)} .
		\label{eq:omega-n-def}
	\end{equation}
	The round hypersphere of radius $R$ is $\SnR=\{R\omega:\omega\in\Sn\}$.
	and we parametrize it by 
	\begin{equation}
		X(\omega)=R\omega \;,
	\end{equation} 
	with $\omega \in S^n$.
	Its surface area element and total area are denoted by
	$\dd\sigma=R^n\dd\Omega$, and $|\SnR|=R^n |\Sn|$, respectively.
	

	\section{Rank-one hyperspherical delta interaction}
	\label{Sec2:Rank-one hyperspherical delta interaction}
	Let $d=n+1$ and let
	\begin{equation}
		H_0=-\Delta,
	\end{equation}
	where the domain is $\mathcal D(H_0)=H^2(\RR^d)$ and we use units $\hbar=2m=1$. The formal Hamiltonian representing the rank one hyperspherical delta interaction is given by
	\begin{equation}
		H=H_0-\lambda\ketbra{\delta_{\SnR}}{\delta_{\SnR}},
		\qquad \lambda>0.
		\label{eq:formal-H}
	\end{equation}
	Here the normalized hyperspherical distribution is
	defined by
	\begin{equation}
		\langle \delta_{\SnR}|\psi\rangle
		=\frac{1}{|S^{n}_R|}\int_{\SnR}\psi(x)\dd\sigma(x)
		=\frac{1}{|\Sn|}\int_{\Sn}\psi(R\omega)\dd\Omega_n(\omega).
		\label{eq:normalized-sphere-state}
	\end{equation}
	Here \(x\) denotes a point on the hypersphere
	\(S^n_{R}\subset\mathbb{R}^{n+1}\). Then, it is well-known that the resolvent associated with the above formal Hamiltonian (\ref{eq:formal-H}) is given by Krein's formula (\ref{eq:Kreinsphere}). For negative energy we write $E=-\nu^2$, $\nu>0$. Since the bound states are the poles of the resolvent, the zeroes of
	the function $\Phi(-\nu^2)$, given by (\ref{eq:principal}), gives the bound state energies.
	
	\begin{remark}
		The distribution $\ket{\delta_{\SnR}}$ is a hypersurface measure of codimension one.  The
		matrix element $\langle \delta_{\SnR}|R_0(-\nu^2)|\delta_{\SnR}\rangle$ is finite for every
		$n\ge1$.  Hence no coupling-constant renormalization is needed for the
		hyperspherical rank-one interaction.  This contrasts with point interactions in
		two and three dimensions, where the diagonal free Green's function (integral kernel of the free resolvent) is divergent.
	\end{remark}
	It is convenient to write the matrix element in the definition of the function $\Phi$ as  
	\begin{equation}
		\langle \delta_{\SnR}|R_0(-\nu^2)|\delta_{\SnR}\rangle
		=\int_{\RR^{n+1}}\frac{|\langle \delta_{\SnR}|p\rangle|^2}{|p|^2+\nu^2}
		\frac{\dd^{n+1}p}{(2\pi)^{n+1}}.
		\label{eq:Gn-def}
	\end{equation}
	where $|p|=\sqrt{p_1^2 + \cdots +p_{n+1}^2}$. It follows from the definition (\ref{eq:normalized-sphere-state}) 
	\begin{align}
		\langle \delta_{\SnR}|p\rangle
		&=\frac{1}{|\Sn|}\int_{\Sn}\e^{\ii p\cdot R\omega}\dd\Omega(\omega).
		\label{eq:sphere-p-state}
	\end{align}
	Since the integral is rotationally invariant, we denote $\langle \delta_{\SnR}|p\rangle$ by $F_n(|p|R)$. 
	It can be computed explicitly using the recursive decomposition 
	
	\begin{eqnarray}
		\dd\Omega_n
		=  \sin^{n-1}\theta \, \dd \theta
		\left[
		\prod_{j=2}^{n-1}
		\left(\sin^{\,n-j}\theta_j\,\dd\theta_j\right)
		\dd \theta_n
		\right] =\sin^{n-1}\theta \,\dd\theta \dd \Omega_{n-1}
	\end{eqnarray}
	so that we obtain 
	\begin{align}
		F_n(|p|R)
		&=\frac{|S^{n-1}|}{|S^n|}
		\int_0^\pi \e^{\ii |p|R\cos\theta}\sin^{n-1}\theta\dd\theta
		\label{eq:sphere-average-fourier2}
	\end{align}
	where $\theta$ is the angle between $p$ and $\omega$. Using the following integral representation of the 
	Bessel function (formula 8.411.4 in \cite{GradshteynRyzhik})
	\begin{align}
		J_m(z)
		&=2\frac{(z/2)^m}{\sqrt\pi\,\Gamma(m+1/2)}
		\int_0^{\pi/2}\cos(z\cos\theta)\sin^{2m}\theta\dd\theta \nonumber \\
		&=\frac{(z/2)^m}{\sqrt\pi\,\Gamma(m+1/2)}
		\int_0^{\pi}\cos(z\cos\theta)\sin^{2m}\theta\dd\theta,
	\end{align}
	and the fact that the imaginary part of $\e^{\ii |p|R\cos\theta}$ integrates to zero by the symmetry
	$\theta\mapsto\pi-\theta$, we get
	\begin{align}
		F_n(|p|R)
		=\Gamma\!\left(\frac{n+1}{2}\right)
		\left(\frac{2}{|p|R}\right)^{(n-1)/2}
		J_{(n-1)/2}(|p|R) \;. \label{eq:normalizedexponentialintegral}
	\end{align}
	After the integration over the angular part, we obtain
	\begin{equation}
		\langle \delta_{\SnR}|R_0(-\nu^2)|\delta_{\SnR}\rangle    =\frac{\Gamma\!\left(\frac{n+1}{2}\right)}{2\pi^{(n+1)/2}R^{n-1}}
		\int_0^\infty \frac{r J_{(n-1)/2}^2(r R)}{r^2+\nu^2}\dd r \;,
	\end{equation}
	where $r:=|p|$. Using the standard integral (formula 6.535 in \cite{GradshteynRyzhik})
	\begin{equation}
		\int_0^\infty\frac{r J_\alpha^2(r R)}{r^2+\nu^2}\dd r
		=I_\alpha(\nu R)K_\alpha(\nu R) \;, \label{eq:Besselintegral}
	\end{equation}
	for $ \alpha>-1$, we obtain the following result.
	\begin{proposition}
		Let $\alpha=\frac{n-1}{2}$, then
		\begin{equation}
			\Phi(-\nu^2)
			=\frac1\lambda-
			\frac{1}{|S^n|R^{n-1}}I_\alpha(\nu R)K_\alpha(\nu R).
			\label{eq:principal-sphere-n}
		\end{equation}
		The bound-state energies $E_B$ are given by the zeros of the following equation
		\begin{equation}
			\frac1\lambda=
			\frac{1}{|S^n|R^{n-1}}I_{(n-1)/2}(\nu R)K_{(n-1)/2}(\nu R),
			\label{eq:implicit-n}
		\end{equation}
		and $E_B=-\nu^2$.
	\end{proposition}
	The bound-state equation obtained here coincides with the zero-angular-momentum
	sector of the usual hyperspherical delta-shell problem after a simple
	normalization of the coupling constant.  Indeed, the local shell interaction
	$-\sigma\delta(r-R)$ acts in all angular momentum sectors, whereas the rank-one
	interaction $-\lambda|\delta_{\SnR}\rangle\langle \delta_{\SnR}|$ acts only on the constant spherical
	harmonic.  For radial wave functions, $\langle \delta_{\SnR}|\psi\rangle=\psi(R)$, and the
	rank-one potential is equivalent to a local shell interaction with effective
	strength $\sigma=\frac{\lambda}{|\SnR|}$.
	Multiplying \eqref{eq:implicit-n} by $|\SnR|=R^n |\Sn|$ gives
	$\frac1\sigma=R I_\alpha(\nu R)K_\alpha(\nu R)$, which is the zero-angular-momentum delta-shell equation obtained by the
	standard radial matching method \cite{DemiralpBeker2003,Demiralp2005}.  Thus
	the present rank-one model should be understood as the $l=0$ sector of the full
	hyperspherical delta-shell problem, with $\lambda=V_n(R)\sigma$.

	The hyperspherical rank-one bound-state equations are summarized in
	\cref{tab:dimension-comparison}.  We write $z=\nu R$.
	
	\begin{table}[htbp]
		\centering
		\begin{tabular}{c c c c}
			\toprule
			Support & Ambient space & Bound-state equation & Critical coupling \\
			\midrule
			$S^1_R$ & $\RR^2$ & $\displaystyle \frac1\lambda=\frac1{2\pi}I_0(z)K_0(z)$ & $0$ \\ \\
			$S^2_R$ & $\RR^3$ & $\displaystyle \frac1\lambda=\frac1{4\pi R}I_{1/2}(z)K_{1/2}(z)$ & $4\pi R$ \\ \\
			$S^3_R$ & $\RR^4$ & $\displaystyle \frac1\lambda=\frac1{2\pi^2R^2}I_1(z)K_1(z)$ & $4\pi^2R^2$ \\ \\
			$S^4_R$ & $\RR^5$ & $\displaystyle \frac1\lambda=\frac{3}{8\pi^2R^3}I_{3/2}(z)K_{3/2}(z)$ & $8\pi^2R^3$ \\ \\
			$S^n_R$ & $\RR^{n+1}$ & $\displaystyle \frac1\lambda=\frac{\Gamma(\frac{n+1}{2})}{2\pi^{(n+1)/2}R^{n-1}}I_{(n-1)/2}(z)K_{(n-1)/2}(z)$ & $\displaystyle \frac{2(n-1)\pi^{(n+1)/2}R^{n-1}}{\Gamma(\frac{n+1}{2})}$ \\
			\bottomrule
		\end{tabular}
		\caption{Implicit equations for bound states of the rank-one attractive delta interaction supported by $S^n_R\subset\RR^{n+1}$.  For $n=1$ the threshold value is infinite because $I_0(z)K_0(z)$ diverges logarithmically as $z\downarrow0$, so every attractive coupling produces one bound state.  For $n\ge2$, a bound state exists only when $\lambda>\lambda_c^{(n)}(R)$.}
		\label{tab:dimension-comparison}
	\end{table}

	\subsection{Existence and Uniqueness of the Bound State}
	\label{Existence and uniqueness of the bound state}

	The right-hand side of \eqref{eq:implicit-n} can be written as the positive
	integral
	\begin{equation}
		\frac{1}{|S^n|R^{n-1}}I_\alpha(\nu R)K_\alpha(\nu R)
		=\frac{1}{|S^n|R^{n-1}}
		\int_0^\infty\frac{r J_\alpha^2(r R)}{r^2+\nu^2}\dd r\;.
	\end{equation}
	Differentiating under the integral sign gives
	\begin{equation}
		\frac{\partial}{\partial\nu}
		\left[\frac{1}{|S^n|R^{n-1}}I_\alpha(\nu R)K_\alpha(\nu R)\right]
		=-\frac{2\nu}{|S^n|R^{n-1}}
		\int_0^\infty\frac{r J_\alpha^2(r R)}{(r^2+\nu^2)^2}\dd r < 0 \;.
	\end{equation}
	Thus there is at most one bound state in the rank-one hyperspherical sector.
	
	For $n=1$, $\alpha=0$ and
	\begin{equation}
		I_0(z)K_0(z)\sim -\log z,
		\qquad z\downarrow0,
	\end{equation}
	so the right-hand side diverges at threshold and every attractive coupling
	$\lambda>0$ produces one bound state.  For $n\ge2$, $\alpha>0$ and
	\begin{equation}
		I_\alpha(z)K_\alpha(z)\to \frac1{2\alpha}=\frac1{n-1},
	\end{equation}
	as $z\downarrow0$. Hence a bound state exists precisely when
	\begin{equation}
		\lambda>\lambda_c^{(n)}(R),
		\qquad
		\lambda_c^{(n)}(R)
		=(n-1)R^{n-1}|S^n|
		=\frac{2(n-1)\pi^{(n+1)/2}R^{n-1}}
		{\Gamma\!\left(\frac{n+1}{2}\right)}.
		\label{eq:critical-coupling}
	\end{equation}
	The case $n=1$ is exceptional and may be regarded as having zero critical
	coupling.
	
	\section{Bound State Analysis of Rank-One Delta Potential Supported on Deformed Hypersphere}
	\label{Bound State Analysis of Delta Potential Supported on Deformed Hypersphere}
	
	\subsection{Small Normal Deformations}
	\label{Sec3:deformations}

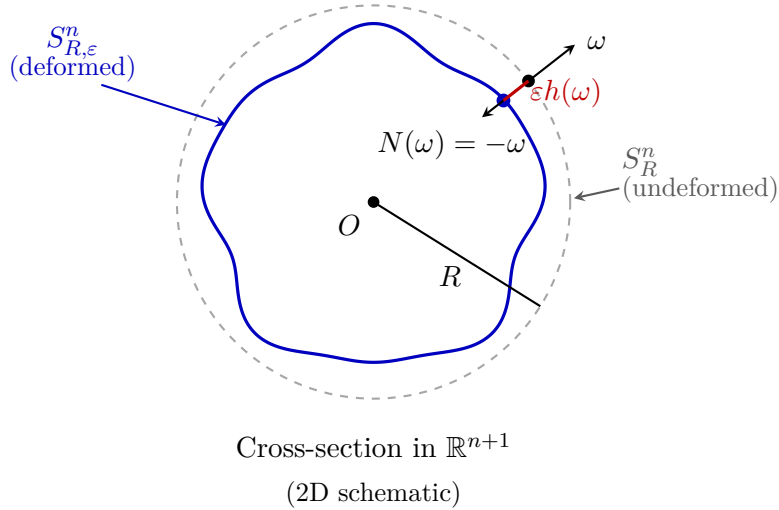
\begin{figure}[ht]
		\centering
		
		\def\figscale{1.0}
		
		\begin{tikzpicture}[scale=\figscale, transform shape, >=stealth]
			
			\def\R{2.6}
			\def\angDef{38}      
			\def\angRad{-32}     
			\def\epsh{0.42}      
			
			\coordinate (O) at (0,0);
			
			\draw[dashed, gray!70, thick] (O) circle (\R);
			
			\draw[blue!75!black, very thick, smooth, samples=220, domain=0:360]
			plot ({((\R-0.42)+0.12*sin(5*\x)+0.06*cos(8*\x))*cos(\x)},
			{((\R-0.42)+0.12*sin(5*\x)+0.06*cos(8*\x))*sin(\x)});
			
			\coordinate (PR) at ({\R*cos(\angRad)},{\R*sin(\angRad)});
			\draw[thick] (O) -- (PR);
			\node at ({0.55*\R*cos(\angRad)-0.20},{0.55*\R*sin(\angRad)-0.22}) {$R$};
			
			\fill (O) circle (2.2pt);
			\node[below left=1pt] at (O) {$O$};
			
			\coordinate (P) at ({\R*cos(\angDef)},{\R*sin(\angDef)});
			
			\coordinate (Q) at ({(\R-\epsh)*cos(\angDef)},{(\R-\epsh)*sin(\angDef)});
			
			\fill[black] (P) circle (2.4pt);
			\fill[blue!75!black] (Q) circle (2.6pt);
			
			\draw[->, thick] (P) -- ++({0.78*cos(\angDef)},{0.78*sin(\angDef)});
			\node[right] at ({(\R+0.82)*cos(\angDef)},{(\R+0.82)*sin(\angDef)}) {$\omega$};
			
			\draw[->, thick] (P) -- ++({-0.78*cos(\angDef)},{-0.78*sin(\angDef)});
			\node[above right] at ({(\R-2.2)*cos(\angDef)-0.40},
			{(\R-2.2)*sin(\angDef)+0.18})
			{$N(\omega)=-\omega$};
			
			\draw[red!75!black, very thick] (P) -- (Q);
			\node[red!75!black, right] at ({(\R-0.22)*cos(\angDef)+0.06},
			{(\R-0.22)*sin(\angDef)-0.02})
			{$\varepsilon h(\omega)$};
			
			\node[gray!70!black, right, align=left] at (3.15,0.35)
			{$S_R^n$\\[-1mm]\small (undeformed)};
			
			\node[blue!75!black, left, align=center] at (-3.0,1.95)
			{$S_{R,\varepsilon}^n$\\[-1mm]\small (deformed)};
			
			\draw[->, thick, gray!70!black] (3.25,0.18) -- (2.65,0.05);
			\draw[->, thick, blue!75!black] (-3.60, 1.55) -- (-1.95,1.02);
			
			\node[align=center] at (0,-3.55)
			{Cross-section in $\mathbb{R}^{n+1}$\\[1mm]\small (2D schematic)};
			
		\end{tikzpicture}
		\caption{Schematic illustration of a small normal deformation of the
			hypersphere \(S_R^n\subset\mathbb{R}^{n+1}\). The undeformed hypersphere is
			represented by the dashed gray curve, while the deformed hypersphere
			\(S_{R,\varepsilon}^n\) is shown by the solid blue curve. The vector
			\(\omega\) denotes the outward unit radial direction, and the inward normal is
			therefore \(N(\omega)=-\omega\). In this convention, the deformation is
			described by $X_\varepsilon(\omega)=R\omega-\varepsilon h(\omega)\omega$.}
		\label{fig:deformed-hypersphere}
	\end{figure}

	Small normal deformation of hypersphere is described by 
	$\SnReps=\{\Xeps(\omega):\omega\in\Sn\}$ and its parametrization is given by 
	\begin{equation}
		X_\varepsilon(\omega)
		=
		(R-\varepsilon h(\omega))\omega
		\;, \label{eq:parametrizationofdeformedsphere}
	\end{equation}
	where $\omega\in S^n$, $h\in C^2(\Sn)$ is real-valued (see Fig. \ref{fig:deformed-hypersphere}). Its surface element is denoted by \(d\sigma_\varepsilon\) and total surface area is denoted by $|\SnReps|$. Here \(\varepsilon\) is sufficiently small
	so that \(X_\varepsilon\) is a regular parametrization of the deformed
	hypersphere.


	The normalized deformed hyperspherical delta distribution is then given by
	\begin{equation}
		\langle \delta_{\SnReps}|\psi\rangle
		=\frac1{|\SnReps|}\int_{\SnReps}\psi(x)\dd\sigma_\varepsilon(x) \;.
		\label{eq:deformed-state-def}
	\end{equation}
	Equivalently, using the parametrization $x=X_{\varepsilon}(\omega)$ with \(X_\varepsilon:S^n\to S^n_{R,\varepsilon}\),
	we may pull the integral back to the unit sphere. Here \(x\) denotes a point on the deformed hypersphere
	\(S^n_{R,\varepsilon}\subset\mathbb{R}^{n+1}\). This can be done explicitly in local coordinates \((\theta^1,\ldots,\theta^n)\) on \(S^n\), and 
	$\omega=\omega(\theta^1,\ldots,\theta^n)$ for simplicity. Then the deformed hypersphere is parametrized by
	\begin{equation}
		X_\varepsilon(\theta^1,\ldots,\theta^n)
		=
		r_\varepsilon(\theta^1,\ldots,\theta^n)\omega(\theta^1,\ldots,\theta^n) \;,
	\end{equation}
	where $r_\varepsilon = R- \varepsilon h$. The surface element is obtained from the metric induced by this parametrization.
	Indeed, the tangent vectors to the deformed hypersphere are
	\begin{equation}
		\partial_i X_\varepsilon
		=
		(\partial_i r_\varepsilon)\omega
		+
		r_\varepsilon \partial_i\omega,
		\qquad i=1,\ldots,n \;,
	\end{equation}
	where $\partial_i$ denotes the partial derivative with respect to $\theta_i$. Since \(\omega\in S^n\), we have $\omega\cdot\partial_i \omega=0$. Hence the cross terms vanish when we compute the scalar product of tangent
	vectors. Therefore the induced metric is
	\begin{align}
		g^{(\varepsilon)}_{ij}
		&=
		\partial_i X_\varepsilon\cdot\partial_j X_\varepsilon \notag\\
		&=
		(\partial_i r_\varepsilon)(\partial_j r_\varepsilon)
		+
		r_\varepsilon^2
		\partial_i\omega\cdot\partial_j\omega .
	\end{align}
	If $\gamma_{ij}=
	\partial_i\omega\cdot\partial_j\omega$
	denotes the standard metric on the unit sphere \(S^n\), then
	\begin{equation}
		g^{(\varepsilon)}_{ij}
		=
		r_\varepsilon^2\gamma_{ij}
		+
		(\partial_i r_\varepsilon)(\partial_j r_\varepsilon).
		\label{eq:deformed-induced-metric}
	\end{equation}
	
	The surface element associated with the metric \(g^{(\varepsilon)}_{ij}\) is then 
	$\sqrt{\det g^{(\varepsilon)}}\,d\theta_1\cdots d\theta_n$. On the other hand, the standard surface element on the unit sphere is
	$d\Omega_n=
	\sqrt{\det\gamma}\,
	d\theta_1\cdots d\theta_n$. Then, the pull-back of the surface element of deformed hypersphere is given by 
	\begin{eqnarray}
		X^{*}_{\varepsilon}(\dd \sigma_{\varepsilon}) = \frac{\sqrt{\det g^{(\varepsilon)}}}{\sqrt{\det \gamma}} \; \dd \Omega_n \;.
	\end{eqnarray}
	It remains to compute this determinant. Writing
	$r_i=\partial_i r_\varepsilon$, equation \eqref{eq:deformed-induced-metric} becomes
	$g^{(\varepsilon)}_{ij}
	=
	r_\varepsilon^2\gamma_{ij}
	+r_i r_j$. Equivalently, in matrix notation,
	$g^{(\varepsilon)}
	=
	r_\varepsilon^2\gamma
	+
	rr^T$. Factoring \(r_\varepsilon^2\), we get
	\begin{equation}
		\det g^{(\varepsilon)}
		=
		r_\varepsilon^{2n}
		\det\left(
		\gamma+\frac{1}{r_\varepsilon^2}rr^T
		\right).
	\end{equation}
	Using the rank-one determinant identity \cite{Meyer2023}, 
	$\det(A+uv^T)=\det A\,\left(1+v^T A^{-1}u\right)$, the determinant can be expressed as 
	\begin{equation}
		\det g^{(\varepsilon)}
		=
		r_\varepsilon^{2n}
		\det\gamma
		\left(
		1+
		\frac{1}{r_\varepsilon^2}r^T\gamma^{-1}r
		\right).
	\end{equation}
	In index notation, $r^T\gamma^{-1}r
	=
	\gamma^{ij}
	(\partial_i r_\varepsilon)
	(\partial_j r_\varepsilon)$, where $(\gamma^{ij})=(\gamma_{ij})^{-1}$. Consequently,
	\begin{equation}
		X_\varepsilon^*(d\sigma_\varepsilon)
		=
		r_\varepsilon^n
		\left(
		1+
		\frac{\gamma^{ij}
			(\partial_i r_\varepsilon)
			(\partial_j r_\varepsilon)}
		{r_\varepsilon^2}
		\right)^{1/2}
		d\Omega_n.
		\label{eq:surface-element-general-radial}
	\end{equation}
	Substituting \(r_\varepsilon=R-\varepsilon h\), we obtain
	\begin{equation}
		X_\varepsilon^*(d\sigma_\varepsilon)
		=
		R^n
		\left(
		1-\frac{n\varepsilon h}{R}
		\right)
		d\Omega_n
		+
		O(\varepsilon^2).
		\label{eq:surface-element-first-order}
	\end{equation}
	where the gradient term does not
	contribute to first order in \(\varepsilon\). Hence
	\begin{eqnarray}
		\langle \delta_{S^n_{R,\varepsilon}}|\psi\rangle
		& = &
		\frac{1}{|\SnReps|}
		\int_{S^n}\psi(X_\varepsilon(\omega))\,
		X_\varepsilon^*(d\sigma_\varepsilon)(\omega) \nonumber \\ & = &  \frac{1}{|\SnReps|}
		\int_{S^n}\psi(X_\varepsilon(\omega))\, R^n
		\left(
		1-\frac{n\varepsilon h(\omega)}{R}
		\right)
		d\Omega_n(\omega) + O(\varepsilon^2)\;.
	\end{eqnarray}

	\subsection{First Order Analysis of Bound States}
	\label{First Order Analysis of Bound States for Delta Potential Supported by Deformed Hypersphere}

	The resolvent associated with the Hamiltonian 
	\begin{equation}
		H=H_0-\lambda\ketbra{\delta_{S^{n}_{R,\varepsilon}}}{\delta_{S^{n}_{R,\varepsilon}}},
		\qquad \lambda>0.
		\label{eq:formal-Hhyper}
	\end{equation}
	is similarly given by the Krein's formula:
	\begin{equation}
		R_\varepsilon(E)=(H_\varepsilon-E)^{-1}
		=R_0(E)+R_0(E)\ket{\delta_{S^{n}_{R,\varepsilon}}} \frac{1}{\Phi_{\varepsilon}(E)} \bra{\delta_{S^{n}_{R,\varepsilon}}}R_0(E),
		\label{eq:resolvent}
	\end{equation}
	where
	\begin{equation}
		\Phi_\varepsilon(E)
		=
		\frac{1}{\lambda}
		-
		\langle \delta_{S^n_{R,\varepsilon}}|
		R_0(E)
		|\delta_{S^n_{R,\varepsilon}}\rangle .
		\label{eq:deformed-principal-function}
	\end{equation}
	Then, the bound state energies $E=-\nu^2$ are given by the zeros of the above function $\Phi_{\varepsilon}(E)$.
	We now compute the matrix element $\langle \delta_{S^n_{R,\varepsilon}}|
	R_0(-\nu^2)
	|\delta_{S^n_{R,\varepsilon}}\rangle$ up to first-order in $\varepsilon$.  Expanding
	\begin{equation}
		e^{i p\cdot X_\varepsilon(\omega)}
		=
		e^{i p\cdot R\omega}
		\left[
		1-i\varepsilon h(\omega)\,p\cdot\omega
		\right]
		+
		O(\varepsilon^2) \;,
	\end{equation}
	and
	\begin{equation}
		|S^{n}_{R, \varepsilon}| = R^n |S^n| \left(1- \frac{n \varepsilon}{R} \langle h \rangle \right) + O(\varepsilon^2)
	\end{equation}
	where $\langle h \rangle := \frac{1}{|S^n|} \int_{S^n} h(\omega) \dd \Omega(\omega)$, we obtain
	\begin{eqnarray}
		& & \langle \delta_{S^n_{R,\varepsilon}}|p \rangle
		=  
		\frac{1}{|\SnReps|}
		\int_{S^n}
		e^{i(p \cdot X_\varepsilon(\omega))}\,
		R^n
		\left(
		1-\frac{n\varepsilon h(\omega)}{R}
		\right)
		d\Omega_n(\omega) + O(\varepsilon^2), \nonumber \\ & & = \frac{\left(1+ \frac{n \varepsilon \langle h \rangle}{R}\right)}{|\Sn|}
		\int_{S^n}
		\left(e^{i p\cdot R\omega}
		\left[
		1-i\varepsilon h(\omega)\,p\cdot\omega
		\right] \right)  \left(
		1-\frac{n\varepsilon h(\omega)}{R}
		\right)
		d\Omega_n(\omega) + O(\varepsilon^2)
		\label{eq:deformed-momentumdistribution-pullback}
	\end{eqnarray}
	For simplicity, let us define $F_{n,\varepsilon}(|p|R)= \langle \delta_{S^n_{R,\varepsilon}}|p \rangle$ and split it as 
	\begin{equation}
		F_{n,\varepsilon}(|p| R)
		=
		F^{(0)}(|p|R)+ \varepsilon F^{(1)}(|p| R) +O(\varepsilon^2),
		\label{eq:Feps-expansion}
	\end{equation}
	where the zeroth order term in $\varepsilon$ is radially symmetric and real-valued 
	\begin{equation}
		F^{(0)}(|p|R)
		:= F_n(|p|R) = 
		\frac{1}{|S^n|}\int_{S^n}e^{i p\cdot R\omega}\,d\Omega_n(\omega) := \frac{S_0(|p|R)}{|S^n|} ,
		\label{eq:F0-S0-def}
	\end{equation}
	and the first order term in $\varepsilon$ is given by
	\begin{equation}
		F^{(1)}(|p|R)
		=
		\frac{1}{|S^n|}
		\left[
		\frac{n\langle h\rangle}{R}S_0(|p|R)
		-
		\frac{n}{R}S_h(|p|R)
		-
		\partial_R S_h(|p|R)
		\right].
		\label{eq:F1-def}
	\end{equation}
	Here
	\begin{equation}
		S_h(|p|R)
		:=
		\int_{S^n}
		h(\omega)e^{i p\cdot R\omega}\,d\Omega_n(\omega).
		\label{eq:Sh-def}
	\end{equation}
	The last term in \eqref{eq:F1-def} follows from the identity
	\begin{equation}
		\partial_R S_h(|p|R)
		=
		\int_{S^n}
		i(p\cdot\omega)h(\omega)e^{i p\cdot R\omega}\,
		d\Omega_n(\omega).
	\end{equation}
	The resolvent matrix element can be then written as
	\begin{eqnarray}
		& & \langle \delta_{S^n_{R,\varepsilon}}|
		R_0(-\nu^2)
		|\delta_{S^n_{R,\varepsilon}}\rangle
		=
		\int_{\mathbb R^{n+1}}
		\frac{|F_{n,\varepsilon}(|p|R)|^2}{p^2+\nu^2}
		\frac{d^{n+1}p}{(2\pi)^{n+1}} \nonumber 
		\\ & & = \int_{\mathbb R^{n+1}}
		\frac{|F^{(0)}|^2}{p^2+\nu^2}
		\frac{d^{n+1}p}{(2\pi)^{n+1}} + \varepsilon  \int_{\mathbb R^{n+1}}
		\frac{\left(F^{(0)} \overline{F^{(1)}} + F^{(1)} \overline{F^{(0)}} \right)}{p^2+\nu^2}
		\frac{d^{n+1}p}{(2\pi)^{n+1}} +  O(\varepsilon^2)\;, \label{eq:matrix-element-form-factor}
	\end{eqnarray}
	We now show how the angular integrations in $F_{n,\varepsilon}(|p|R)$ are evaluated. 
	For \(n\geq 2\),
	we use the standard plane-wave expansion in hyperspherical harmonics
	\cite{Avery1989,WenAvery1985}.  Let
	$\alpha=\frac{n-1}{2}$, and $\widehat p=\frac{p}{|p|}$, then
	\begin{equation}
		e^{i p\cdot R\omega}
		=
		(2\pi)^{(n+1)/2}
		\sum_{\ell=0}^{\infty}
		\sum_{m=0}^{d_\ell-1}
		i^\ell
		\frac{J_{\ell+\alpha}(|p|R)}
		{(|p|R)^\alpha}
		Y_{\ell m}(\omega)
		\overline{Y_{\ell m}(\widehat p)} \;.
		\label{eq:plane-wave-hyperspherical-expansion}
	\end{equation}
	Here \(d_\ell = \frac{(2 \ell +n -1) (\ell +n -2)!}{\ell ! (n-1)!}\) is the dimension of the space of hyperspherical harmonics of
	degree \(\ell\) on \(S^n\), and the harmonics are normalized by
	\begin{eqnarray}
		\int_{S^n}
		\overline{Y_{\ell m}(\omega)}
		Y_{\ell' m'}(\omega)\,d\Omega_n(\omega)
		=
		\delta_{\ell\ell'}\delta_{mm'}.
	\end{eqnarray}
	The zeroth harmonic is constant $Y_{00}(\omega)=\frac{1}{\sqrt{|S^n|}}$,
	and $d_0=1$. Here the index $m$ should not be confused with the usual magnetic quantum number in $S^2$.

	The angular integration over \(\widehat{p}\)
	selects the \(\ell=0\) and $m=0$ terms in the order $\varepsilon$ term in equation (\ref{eq:matrix-element-form-factor}) thanks to
	\begin{eqnarray}
		\int_{S^n}
		\overline{Y_{\ell m}(\widehat{p})} \,d\Omega_{n}(\widehat{p})
		=
		\sqrt{|S^n|} \delta_{\ell 0}\delta_{m0} \;,
	\end{eqnarray}
	where $\dd \Omega_n(\widehat{p})$ is the surface element of hypersphere in momentum space. Then, 
	\begin{eqnarray}
		& & \int_{S^n}
		S_h(|p|R)\,d\Omega_n(\widehat p)
		=   \int_{S^n} \int_{S^n} (2\pi)^{(n+1)/2}
		\sum_{\ell=0}^{\infty}
		\sum_{m=0}^{d_\ell -1}
		i^\ell
		\frac{J_{\ell+\alpha}(|p|R)}
		{(|p|R)^\alpha} \nonumber \\ & & \hspace{7cm} \times 
		Y_{\ell m}(\omega) h(\omega)
		\overline{Y_{\ell m}(\widehat p)} \; \dd \Omega(\omega) \dd \Omega(\widehat{p})\nonumber \\ & & \hspace{3.7cm} = \sqrt{|S^n|}\int_{S^n} (2\pi)^{(n+1)/2}
		\frac{J_{\alpha}(|p|R)}
		{(|p|R)^\alpha} Y_{00}(\omega) h(\omega)
		\; \dd \Omega(\omega) \nonumber \\ &  & \hspace{3.7cm}=   
		|S^n|\langle h\rangle S_0(|p| R),
		\label{eq:intSh}
	\end{eqnarray}
	and similarly
	\begin{eqnarray}
		& &  \int_{S^n}
		\partial_R S_h(|p|R)\,d\Omega_n(\widehat p) = \int_{S^n} \int_{S^n} (2\pi)^{(n+1)/2}
		\sum_{\ell=0}^{\infty}
		\sum_{m=0}^{d_\ell -1}
		i^\ell
		\partial_R \left(\frac{J_{\ell+\alpha}(|p|R)}
		{(|p|R)^\alpha} \right) \nonumber \\ & & \hspace{7cm} \times 
		Y_{\ell m}(\omega) h(\omega)
		\overline{Y_{\ell m}(\widehat p)} \; \dd \Omega(\omega) \dd \Omega(\widehat{p})\nonumber \\ & & \hspace{3.7cm} = \sqrt{|S^n|}\int_{S^n} (2\pi)^{(n+1)/2}
		\partial_R \left(\frac{J_{\alpha}(|p|R)}
		{(|p|R)^\alpha} \right) Y_{00}(\omega) h(\omega)
		\; \dd \Omega(\omega) \nonumber \\ &  & \hspace{3.7cm}=   
		|S^n|\langle h\rangle \partial_R S_0(|p|R),
		\label{eq:angular-average-dR-Sh}
	\end{eqnarray}
	Combining all these results, we obtain
	\begin{eqnarray}
		& & \int_{\mathbb R^{n+1}}
		\frac{\left(F^{(0)} \overline{F^{(1)}} + F^{(1)} \overline{F^{(0)}} \right)}{p^2+\nu^2}
		\frac{d^{n+1}p}{(2\pi)^{n+1}} \nonumber \\ & & \hspace{4cm}= - \frac{2 \langle h \rangle}{(2\pi)^{n+1}|S^n|} \int_{0}^{\infty} \frac{r^n}{r^2 + \nu^2}  \left( S_0(r R) \partial_R S_0(r R)\right)\dd r \;,
	\end{eqnarray}
	where $r=|p|$. The expression inside the parentheses can be computed as
	\begin{eqnarray}
		S_0(r R) \partial_R S_0(r R) & = &  (2\pi)^{n+1} \frac{J_{\frac{n-1}{2}}(r R)}
		{(r R)^{\frac{n-1}{2}}} \partial_R \left(\frac{J_{\frac{n-1}{2}}(r R)} 
		{(r R)^{\frac{n-1}{2}}}  \right) \nonumber \\ & = & - \frac{(2\pi)^{n+1} r^2 R}{(r R)^n} J_{\frac{n-1}{2}}(r R) J_{\frac{n+1}{2}}(r R) \;.
	\end{eqnarray}
	Here we have used the identity $\frac{d}{dz}J_\alpha(z)=\frac{\alpha}{z} J_\alpha(z)-J_{\alpha+1}(z)$ \cite{LebedevSpecialFunctions}. Then,
	\begin{eqnarray}
		& & \int_{\mathbb R^{n+1}}
		\frac{\left(F^{(0)} \overline{F^{(1)}} + F^{(1)} \overline{F^{(0)}} \right)}{p^2+\nu^2}
		\frac{d^{n+1}p}{(2\pi)^{n+1}} \nonumber \\ & & \hspace{4cm}= \frac{2 \langle h \rangle}{|S^n|R^{n-1}} \int_{0}^{\infty} \frac{r^2}{r^2 + \nu^2}  J_{\frac{n-1}{2}}(r R) J_{\frac{n+1}{2}}(r R)  \, \dd r \;.
	\end{eqnarray}
	Decomposing $\frac{r^2}{r^2 + \nu^2} = 1-\frac{\nu^2}{r^2+\nu^2}$ and using the symmetry of the momentum volume element under $p\to-p$ and the formulas (6.512) and (6.577) in \cite{GradshteynRyzhik},  we can evaluate the integrals 
	\begin{eqnarray}
		& & \int_{\mathbb R^{n+1}}
		\frac{\left(F^{(0)} \overline{F^{(1)}} + F^{(1)} \overline{F^{(0)}} \right)}{p^2+\nu^2}
		\frac{d^{n+1}p}{(2\pi)^{n+1}} \nonumber \\ & & \hspace{4cm}=  \frac{\langle h \rangle}{|S^n|R^{n-1}} \left(\frac{1}{R}-2\nu I_{\frac{n+1}{2}}(\nu R) K_{\frac{n-1}{2}}(\nu R)\right)\;.
	\end{eqnarray}
	Using the identity $I_{\alpha+1}(z)K_\alpha(z)
	+I_\alpha(z)K_{\alpha+1}(z)
	=\frac{1}{z}$ \cite{LebedevSpecialFunctions}, we finally obtain
	\begin{align}
		\langle \delta_{S^n_{R,\varepsilon}}|
		R_0(-\nu^2)
		|\delta_{S^n_{R,\varepsilon}}\rangle
		&=
		\frac{1}{|S^n|R^{n-1}}
		I_{\frac{n-1}{2}}(\nu R)K_{\frac{n-1}{2}}(\nu R)
		\notag\\
		&\quad
		+
		\varepsilon
		\frac{\langle h\rangle}{|S^n|R^{n-1}}
		\left[
		-\frac{1}{R}
		+
		2\nu I_{\frac{n-1}{2}}(\nu R)K_{\frac{n+1}{2}}(\nu R)
		\right]
		+
		O(\varepsilon^2).
		\label{eq:matrix-element-final-harmonic}
	\end{align}
	Therefore 
	\begin{align}
		\Phi_\varepsilon(-\nu^2)
		&=
		\frac{1}{\lambda}
		-
		\frac{1}{|S^n|R^{n-1}}
		I_{\frac{n-1}{2}}(\nu R)K_{\frac{n-1}{2}}(\nu R)
		\notag\\
		&\quad
		-
		\varepsilon
		\frac{\langle h\rangle}{|S^n|R^{n-1}}
		\left[
		-\frac{1}{R}
		+
		2\nu I_{\frac{n-1}{2}}(\nu R)K_{\frac{n+1}{2}}(\nu R)
		\right]
		+
		O(\varepsilon^2).
		\label{eq:principal-final-harmonic}
	\end{align}
	
	Hence, let us now summarize the result as 
	\begin{theorem}[Bound-state equation for the deformed hypersphere]
		Let
		\[
		S^n_{R,\varepsilon}
		=
		\left\{
		X_\varepsilon(\omega)
		=
		(R-\varepsilon h(\omega))\omega
		:
		\omega\in S^n
		\right\}
		\subset \mathbb{R}^{n+1},
		\]
		where \(h\in C^2(S^n)\), and let
		\[
		H_\varepsilon
		=
		H_0
		-
		\lambda
		\ketbra{\delta_{S^n_{R,\varepsilon}}}
		{\delta_{S^n_{R,\varepsilon}}},
		\qquad \lambda>0,
		\]
		where \(\delta_{S^n_{R,\varepsilon}}\) denotes the normalized hyperspherical
		delta distribution supported on \(S^n_{R,\varepsilon}\).  Let $\langle h\rangle$ be the average of $h$ over $S^n$.
		Then, up to first order in \(\varepsilon\), the negative bound-state energies
		\(E=-\nu^2\), \(\nu>0\), are determined by the zeros of the function
		\[
		\Phi_\varepsilon(-\nu^2)=0,
		\]
		where
		\begin{align}
			\Phi_\varepsilon(-\nu^2)
			&=
			\frac{1}{\lambda}
			-
			\frac{1}{|S^n|R^{n-1}}
			I_{\frac{n-1}{2}}(\nu R)
			K_{\frac{n-1}{2}}(\nu R)
			\notag\\
			&\quad
			-
			\varepsilon
			\frac{\langle h\rangle}{|S^n|R^{n-1}}
			\left[
			-\frac{1}{R}
			+
			2\nu
			I_{\frac{n-1}{2}}(\nu R)
			K_{\frac{n+1}{2}}(\nu R)
			\right]
			+
			O(\varepsilon^2).
			\label{eq:theorem-deformed-principal-function}
		\end{align}
	\end{theorem}
	
	\begin{remark}
		There is an alternative short derivation for the above integrals.
The angular integrations in (3.28) and (3.29) can also be evaluated without using the plane-wave expansion. For fixed $\omega\in S^n$ and $r=|p|$, rotational invariance of the surface measure gives
		\[
		\int_{S^n}e^{irR\widehat p\cdot\omega}\,\dd\Omega_n(\widehat p)=S_0(rR),
		\]
		independently of $\omega$. Interchanging the integrations in the definition (3.22), we immediately obtain
		\begin{align*}
			\int_{S^n}S_h(r R)\,\dd\Omega_n(\widehat p)
			&=\int_{S^n}h(\omega)\left[\int_{S^n}e^{irR\widehat p\cdot\omega}\,\dd\Omega_n(\widehat p)\right]\dd\Omega_n(\omega)\\
			&=|S^n|\hs S_0(rR).
		\end{align*}
		Differentiation with respect to $R$ similarly yields
		\[
		\int_{S^n}\partial_R S_h(r R)\,\dd\Omega_n(\widehat p)
		=|S^n|\hs\partial_R S_0(rR).
		\]
		These are exactly (3.28) and (3.29), and substituting them into (3.21) reproduces (3.30). In addition, the first variation of the resolvent matrix element is linear in $h$ and invariant under rotations of $h$. Hence it depends only on $\hs$. For the constant deformation $h(\omega)=c$, the deformed support is a round hypersphere of radius $R-\varepsilon c$, and therefore
		\[
		\langle\delta_{S_{R,\varepsilon}^n}|R_0(-\nu^2)|\delta_{S_{R,\varepsilon}^n}\rangle
		=M_0(\nu,R)-\varepsilon\hs\,\partial_R M_0(\nu,R)+O(\varepsilon^2),
		\qquad M_0(\nu,R)=\frac{I_\alpha(\nu R)K_\alpha(\nu R)}{|S^n|R^{n-1}},
		\]
		where $\alpha=(n-1)/2$. The Bessel derivative identities give
		\[
		\partial_RM_0(\nu,R)=\frac{1}{|S^n|R^{n-1}}\left[\frac1R-2\nu I_\alpha(\nu R)K_{\alpha+1}(\nu R)\right],
		\]
		which independently confirms the corrected signs in (3.34)--(3.36) and the effective-radius interpretation.
	\end{remark}

	\begin{remark}
		For \(n=1\), the hyperspherical expansion
		\eqref{eq:plane-wave-hyperspherical-expansion} should be replaced by the
		ordinary Fourier--Bessel expansion on the circle.  Indeed, in this case
		\(\alpha=(n-1)/2=0\), and the Gegenbauer-polynomial derivation of
		\eqref{eq:plane-wave-hyperspherical-expansion} becomes singular.  Writing
		$\omega=(\cos\theta,\sin\theta)$,
		$\widehat p=(\cos\theta_p,\sin\theta_p)$, we have
		$ p\cdot R\omega
		=
		|p|R\cos(\theta-\theta_p)$. Therefore one uses the Jacobi--Anger expansion
		\begin{equation}
			e^{i|p|R\cos(\theta-\theta_p)}
			=
			\sum_{k=-\infty}^{\infty}
			i^kJ_k(|p|R)e^{ik(\theta-\theta_p)}.
			\label{eq:jacobi-anger-complex}
		\end{equation}
		Equivalently, in the normalized Fourier basis
		\[
		Y_k(\theta)=\frac{1}{\sqrt{2\pi}}e^{ik\theta},
		\qquad k\in\mathbb Z,
		\]
		this can be written as
		\begin{equation}
			e^{i|p|R\cos(\theta-\theta_p)}
			=
			2\pi
			\sum_{k=-\infty}^{\infty}
			i^kJ_k(|p|R)
			Y_k(\theta)\overline{Y_k(\theta_p)}.
			\label{eq:s1-plane-wave-expansion}
		\end{equation}
		The real form is
		\begin{equation}
			e^{i|p|R\cos\theta}
			=
			J_0(|p|R)
			+
			2\sum_{m=1}^{\infty}
			i^mJ_m(|p|R)\cos(m\theta).
			\label{eq:jacobi-anger-real}
		\end{equation}
		Thus the \(n=1\) case is fully consistent with the higher-dimensional
		calculation, but it must be treated by the standard Fourier basis on \(S^1\)
		rather than by the Gegenbauer formulation.
	\end{remark}

	\begin{remark}
		In the case of deformed circle problem, which corresponds to $n=1$, $\alpha=0$, and $|S^1|=2\pi$, we get
		\begin{equation}
			\Pheps(-\nu^2)
			=\frac1\lambda-\frac1{2\pi}I_0(\nu R)K_0(\nu R)
			-\frac{\varepsilon}{2\pi^2}
			\left(-\frac1{2R}+\nu I_0(\nu R)K_1(\nu R)\right)
			\int_0^{2\pi}h(\theta)\dd\theta+O(\varepsilon^2) \;.
			\label{eq:circle-analog-419}
		\end{equation}
		In the case of deformed sphere problem which corresponds to $n=2$, $\alpha=1/2$, and $|S^2|=4\pi$, we get
		\begin{align}
			\Pheps(-\nu^2)
			&=\frac1\lambda-\frac1{4\pi R}I_{1/2}(\nu R)K_{1/2}(\nu R) \nonumber \\
			&\quad -
			\frac{\varepsilon}{8\pi^2R}
			\left(-\frac1{2R}+\nu I_{1/2}(\nu R)K_{3/2}(\nu R)\right)
			\int_{S^2}h(\omega)\dd\omega+O(\varepsilon^2).
			\label{eq:sphere-analog-517}
		\end{align}
		These are the exactly the same formulas for the circular and spherical first-order implicit equations obtained in \cite{ErmanSeymenTurgut2022}. 
	\end{remark}

	Since the function $\Phi_{\varepsilon}$ has been computed up to order \(\varepsilon\), the
	corresponding bound-state energy can also be determined to the same order.
	Let
	\[
	\nu=\nu_0+\varepsilon\nu_1+O(\varepsilon^2),
	\]
	where \(\nu_0\) is the solution of the undeformed hyperspherical problem.  The
	bound-state energy of the deformed hyperspherical support is then
	\[
	E_B(\varepsilon)=-(\nu_0+\varepsilon\nu_1)^2.
	\]
	Substituting this expansion into the first-order function $\Phi_{\varepsilon}$, and expanding the Bessel functions around $\varepsilon=0$:
	\begin{eqnarray}
		I_{\alpha}\left( (\nu_0+\varepsilon \nu_1)R\right) & = &  I_{\alpha}\left(\nu_0 R\right) + \varepsilon \nu_1 R \left( \frac{\alpha}{\nu_0 R} I_{\alpha}(\nu_0 R) + I_{\alpha+1}(\nu_0 R)\right) + O(\varepsilon^2) \\    K_{\alpha}\left( (\nu_0+\varepsilon \nu_1 )R\right) & = &  K_{\alpha}\left(\nu_0 R\right) + \varepsilon \nu_1 R \left( \frac{\alpha}{\nu_0 R} K_{\alpha}(\nu_0 R) - K_{\alpha+1}(\nu_0 R)\right) + O(\varepsilon^2)
	\end{eqnarray}
	the equation $\Phi_{\varepsilon}(-(\nu_0+\varepsilon\nu_1)^2)=0$ yields
	\begin{equation}
		\nu_1 = \frac{\langle h \rangle}{R} \left(\frac{\frac{1}{R}-2 \nu_0 I_{\alpha} (\nu_0 R) K_{\alpha+1}(\nu_0 R)}{\frac{2\alpha}{\nu_0 R}I_{\alpha}(\nu_0 R) K_{\alpha}(\nu_0 R)+K_{\alpha}(\nu_0 R) I_{\alpha+1}(\nu_0 R) - I_{\alpha}(\nu_0 R)K_{\alpha+1}(\nu_0 R)} \right) \;. \label{eq:nu1}
	\end{equation}
	Here, the zeroth-order part cancels by the unperturbed bound-state equation.  Solving
	the remaining first-order equation for \(\nu_1\) gives the above explicit expression for
	the first-order correction $\nu_1$.  Consequently,
	\begin{eqnarray}
		& &  E_B(\varepsilon)
		\nonumber \\ & & =
		-\nu_0^2-
		\varepsilon
		\frac{2\nu_0 \langle h \rangle}{R}
		\left(\frac{\frac{1}{R}-2 \nu_0 I_{\alpha} (\nu_0 R) K_{\alpha+1}(\nu_0 R)}{\frac{2\alpha}{\nu_0 R}I_{\alpha}(\nu_0 R) K_{\alpha}(\nu_0 R)+K_{\alpha}(\nu_0 R) I_{\alpha+1}(\nu_0 R) - I_{\alpha}(\nu_0 R)K_{\alpha+1}(\nu_0 R)} \right) 
		\nonumber \\ & & \hspace{5cm}+
		O(\varepsilon^2).
		\label{eq:circle-energy-first-order-unsimplified}
	\end{eqnarray}

	\begin{remark}
		This simple form admits a useful geometric interpretation.  To first order, a
		deformed hyperspherical support behaves, for the rank-one interaction, like a hypersphere
		whose radius is shifted by the average normal deformation.  With the inward
		normal convention, this means replacing the original radius \(R\) by
		\[
		R-\varepsilon\langle h\rangle,
		\]
		the corresponding undeformed hyperspherical bound-state equation is
		\begin{equation}
			\frac{1}{\lambda}
			-
			\frac{1}{|S^n|(R-\varepsilon \langle h \rangle)^{n-1}}
			I_\alpha\!\left((\nu_0+\varepsilon\nu_1)(R-\varepsilon\langle h\rangle)\right)
			K_\alpha\!\left((\nu_0+\varepsilon\nu_1)(R-\varepsilon\langle h\rangle)\right)
			=
			0.
			\label{eq:circle-effective-radius-equation}
		\end{equation}
		Expanding this equation up to first order in \(\varepsilon\), we obtain exactly same $\nu_1$ given by (\ref{eq:nu1}).
	\end{remark}

	\begin{corollary}
		For an attractive rank-one delta interaction supported by a hypersphere, a small
		inward normal deformation changes the bound-state energy, to first order, in
		the same way as replacing the original hypersphere by a hypersphere whose radius is
		$R-\varepsilon\langle h\rangle$. 
	\end{corollary}

	\begin{corollary}[Mean-zero deformations]
		If
		\begin{equation}
			\int_{\Sn}h(\omega)\dd\Omega(\omega)=0,
		\end{equation}
		then the bound-state energy does not change at first order:
		\begin{equation}
			E_B(\varepsilon)=E_B^{(0)}(R)+O(\varepsilon^2).
		\end{equation}
	\end{corollary}

	\begin{remark}
		Several features are apparent.  First, the order of the Bessel functions
		increases by one half whenever the dimension of the support is increased by
		one.  Second, the circle is exceptional because the Bessel product diverges at
		threshold.  Third, in every dimension $n\ge2$ the Bessel product has a finite
		threshold value, so a critical coupling is required.  Finally, the first-order
		deformation formula is dimension-independent in its geometric interpretation:
		replace the radius by $R-\varepsilon\langle h\rangle$ and solve the same
		implicit equation.
	\end{remark}

	\begin{remark}
		We also mention that different notions of deformation for supports of delta
		interactions have been studied in the literature.  For example, in
		\cite{ExnerYoshitomi2003}, asymptotic expansions of eigenvalues were obtained for
		Hamiltonians of the type
		$-\Delta - \beta \delta(\,\cdot-\Sigma\setminus S_\varepsilon)$,
		where \(S_\varepsilon\) denotes a family of measurable subsets of a fixed
		surface \(\Sigma\), with size vanishing as \(O(\varepsilon)\) as
		\(\varepsilon\to 0\).  In that case, the support is modified by deleting a
		small subset from a fixed surface.  The deformation considered in the present
		work is of a different nature: we deform the entire hyperspherical support
		smoothly in the normal direction and study the resulting first-order change in
		the bound-state energy.
	\end{remark}

	\section{Conclusion}
	
	In this paper, we have studied rank-one delta interactions supported by small
	normal deformations of \(n\)-dimensional hyperspheres
	\(S_R^n\subset\mathbb{R}^{n+1}\).  Our main contribution is the bound-state
	analysis of these singular interactions under the assumption that the
	deformation of the support is sufficiently small.  The undeformed rank-one
	interaction supported on \(S_R^n\) provides the reference model, and its
	principal function can be computed explicitly in terms of modified Bessel
	functions.  More precisely, for negative energies \(E=-\nu^2\), the undeformed
	principal function is expressed through the product
	\[
	I_{\frac{n-1}{2}}(\nu R)K_{\frac{n-1}{2}}(\nu R),
	\]
	so that the bound-state energies are determined by a single scalar equation.
	
	This formulation gives a unified description of rank-one hyperspherical delta
	interactions in arbitrary dimension.  It also makes the dependence on the
	dimension transparent.  The circle case, corresponding to \(n=1\), is special:
	there is no critical coupling, and every attractive interaction produces a bound
	state.  For \(n\geq 2\), the threshold value of the Bessel product is finite,
	and a bound state exists only when the coupling constant exceeds the critical
	value \(\lambda_c^{(n)}(R)\).  Thus the hyperspherical formulation clarifies how
	the spectral behavior changes when one passes from the circle in
	\(\mathbb{R}^2\) to higher-dimensional hyperspheres.
	
	The main part of the paper concerns the effect of a small deformation of the
	hyperspherical support in the normal direction.  We derived the first-order
	change of the principal function and, consequently, the first-order shift of the
	bound-state energy.  The result has a particularly simple geometric
	interpretation: to first order in the deformation parameter \(\varepsilon\), the
	bound-state energy depends only on the average normal displacement of the
	support.  Equivalently, for the rank-one interaction, the deformed hypersphere
	is spectrally equivalent up to order \(\varepsilon\) to a round hypersphere
	whose radius is shifted by the average of the normal deformation.  In the inward
	normal convention
	\[
	X_\varepsilon(\omega)
	=
	(R-\varepsilon h(\omega))\omega,
	\qquad \omega\in S^n,
	\]
	this effective radius is
	\[
	R-\varepsilon\langle h\rangle,
	\qquad
	\langle h\rangle
	=
	\frac{1}{|S^n|}
	\int_{S^n}h(\omega)\,d\Omega_n(\omega).
	\]
	
	In particular, deformations with zero average do not affect the rank-one
	bound-state energy at first order.  This reflects the fact that the rank-one
	interaction generated by the normalized hypersurface measure couples only to
	the constant hyperspherical harmonic.  Therefore the nonconstant components of
	the deformation are invisible at leading order.  These components are expected
	to contribute at second order and would also play a more direct role in the
	full delta-shell problem, where non-radial angular momentum sectors are
	present.  The present analysis therefore provides both an explicit solvable
	model for deformed hyperspherical rank-one interactions and a natural starting
	point for studying higher-order deformation effects and more general singular
	interactions supported on hypersurfaces.

	\section*{Appendix: Plane-wave expansion in hyperspherical harmonics}
	\label{app:plane-wave-hyperspherical}
	
	In this appendix we recall the derivation of the plane-wave expansion in
	hyperspherical harmonics.  Let $z=|p|R$,
	$\widehat{p}=\frac{p}{|p|}$. Then
	$e^{i p\cdot R\omega}=
	e^{iz\widehat{p}\cdot\omega}$, $\omega,\widehat{p}\in S^n$.
	
	We assume first that \(n\geq 2\), so that \(\alpha>0\).  Let
	\(\{Y_{\ell m}\}_{m=0}^{d_\ell -1}\) be an orthonormal real basis of hyperspherical
	harmonics of degree \(\ell\) on \(S^n\). For fixed \(\widehat{p}\), the function
	\(\omega\mapsto e^{iz\widehat{p}\cdot\omega}\) has an expansion
	\[
	e^{iz\widehat{p}\cdot\omega}
	=
	\sum_{\ell=0}^{\infty}
	\sum_{m=0}^{d_\ell -1}
	a_{\ell m}(z,\widehat{p})Y_{\ell m}(\omega).
	\]
	By rotational invariance, the coefficients must have the form
	$a_{\ell m}(z,\widehat{p})
	=A_\ell(z)\overline{Y_{\ell m}(\widehat{p})}$.
	Therefore
	\begin{equation}
		e^{iz\widehat{p}\cdot\omega}
		=
		\sum_{\ell=0}^{\infty}
		A_\ell(z)
		\sum_{m=0}^{d_\ell -1}
		Y_{\ell m}(\omega)
		\overline{Y_{\ell m}(\widehat{p})}.
		\label{eq:app-harmonic-expansion-general}
	\end{equation}
	We now use the addition theorem for hyperspherical harmonics \cite{Muller1966,DaiXu2013},
	\begin{equation}
		\sum_{m=0}^{d_\ell -1}
		Y_{\ell m}(\omega)
		\overline{Y_{\ell m}(\omega')}
		=
		\frac{d_\ell}{|S^n|}
		\frac{C_\ell^\alpha(\omega\cdot\omega')}
		{C_\ell^\alpha(1)},
		\label{eq:app-addition-theorem}
	\end{equation}
	where \(C_\ell^\alpha\) is the Gegenbauer polynomial and $C_\ell^\alpha(1)= \binom{\ell + 2\alpha - 1}{\ell} = \frac{\Gamma(\ell + 2\alpha)}{\ell! \, \Gamma(2\alpha)}
	$. Taking \(\omega'=\widehat{p}\), equation
	\eqref{eq:app-harmonic-expansion-general} becomes
	\[
	e^{izt}
	=
	\sum_{\ell=0}^{\infty}
	A_\ell(z)
	\frac{d_\ell}{|S^n|}
	\frac{C_\ell^\alpha(t)}
	{C_\ell^\alpha(1)} \;,
	\qquad
	t=\widehat{p}\cdot\omega.
	\]
	On the other hand, Gegenbauer's expansion of the plane wave is
	\begin{equation}
		e^{izt}
		=
		2^\alpha\Gamma(\alpha)
		\sum_{\ell=0}^{\infty}
		i^\ell(\ell+\alpha)
		\frac{J_{\ell+\alpha}(z)}
		{z^\alpha}
		C_\ell^\alpha(t).
		\label{eq:app-gegenbauer-plane-wave}
	\end{equation}
	Comparing the coefficients of \(C_\ell^\alpha(t)\), we obtain
	\[
	A_\ell(z)
	\frac{d_\ell}{|S^n|C_\ell^\alpha(1)}
	=
	2^\alpha\Gamma(\alpha)
	i^\ell(\ell+\alpha)
	\frac{J_{\ell+\alpha}(z)}
	{z^\alpha}.
	\]
	Using $\frac{d_\ell}{C_\ell^\alpha(1)}
	=
	\frac{\ell+\alpha}{\alpha}$, we get
	\[
	A_\ell(z)
	=
	2^\alpha\Gamma(\alpha+1)|S^n|
	i^\ell
	\frac{J_{\ell+\alpha}(z)}
	{z^\alpha}.
	\]
	Since $|S^n|
	=\frac{2\pi^{\alpha+1}}{\Gamma(\alpha+1)}$, 
	\[
	A_\ell(z)
	=
	(2\pi)^{(n+1)/2}
	i^\ell
	\frac{J_{\ell+\alpha}(z)}
	{z^\alpha}.
	\]
	Substituting this coefficient into
	\eqref{eq:app-harmonic-expansion-general}, we obtain
	\begin{equation}
		e^{iz\widehat{p}\cdot\omega}
		=
		(2\pi)^{(n+1)/2}
		\sum_{\ell=0}^{\infty}
		\sum_{m=0}^{d_\ell-1}
		i^\ell
		\frac{J_{\ell+\alpha}(z)}
		{z^\alpha}
		Y_{\ell m}(\omega)
		\overline{Y_{\ell m}(\widehat{p})}.
		\label{eq:app-plane-wave-z}
	\end{equation}
	Finally, substituting \(z=|p|R\), we arrive at
	\begin{equation}
		e^{i p\cdot R \omega}
		=
		(2\pi)^{(n+1)/2}
		\sum_{\ell=0}^{\infty}
		\sum_{m=0}^{d_\ell-1}
		i^\ell
		\frac{J_{\ell+\alpha}(|p|R)}
		{(|p|R)^\alpha}
		Y_{\ell m}(\omega)
		\overline{Y_{\ell m}(\widehat{p})},
		\qquad
		\alpha=\frac{n-1}{2}.
		\label{eq:app-plane-wave-final}
	\end{equation}
	If a real orthonormal basis of hyperspherical harmonics is used, the complex
	conjugation in \eqref{eq:app-plane-wave-final} may be omitted.
	
	For \(n=1\), the parameter \(\alpha\) vanishes and the Gegenbauer formulation
	has to be replaced by the usual Fourier expansion on \(S^1\).  In that case one
	uses the Jacobi--Anger expansion
	\[
	e^{iz\cos(\theta-\phi)}
	=
	\sum_{k=-\infty}^{\infty}
	i^k J_k(z)e^{ik(\theta-\phi)}.
	\]
	Equivalently, with
	\[
	Y_k(\theta)=\frac{1}{\sqrt{2\pi}}e^{ik\theta},
	\qquad k\in\mathbb Z,
	\]
	one may write
	\[
	e^{iz\cos(\theta-\phi)}
	=
	2\pi
	\sum_{k=-\infty}^{\infty}
	i^k J_k(z)
	Y_k(\theta)\overline{Y_k(\phi)}.
	\]
	Thus \eqref{eq:app-plane-wave-final} is the natural higher-dimensional analogue
	of the standard Jacobi--Anger expansion.

	\section*{DATA AVAILABILITY}
	Data sharing is not applicable to this article as no new data were created or analyzed in this study.

\end{document}